# Thermodynamic potentials and Thermodynamic Relations in Nonextensive Thermodynamics

Guo Lina,  Du Jiulin

*Department of Physics, School of Science, Tianjin University, Tianjin 300072, China*

**Abstract**

The generalized Gibbs free energy and enthalpy is derived in the framework of nonextensive thermodynamics by using the so-called physical temperature and the physical pressure. Some thermodynamical relations are studied by considering the difference between the physical temperature and the inverse of Lagrange multiplier. The thermodynamical relation between the heat capacities at a constant volume and at a constant pressure is obtained using the generalized thermodynamical potential, which is found to be different from the traditional one in Gibbs thermodynamics. But, the expressions for the heat capacities using the generalized thermodynamical potentials are unchanged.

PACS: 05.70.-a; 05.20.-y; 05.90.+m

Keywords: Nonextensive thermodynamics; The generalized thermodynamical relations

**1. Introduction**

In the traditional thermodynamics, there are several fundamental thermodynamic potentials, such as internal energy $U$, Helmholtz free energy $F$, enthalpy $H$, Gibbs free energy $G$. Each of them is a function of temperature $T$, pressure $P$ and volume $V$. They and their thermodynamical relations constitute the basis of classical thermodynamics.

Recently, nonextensive thermo-statitstics has attracted significant interests and has obtained wide applications to so many interesting fields, such as astrophysics [2, 3], real gases [4], plasma [5], nuclear reactions [6] and so on. Especially, one has been studying the problems whether the thermodynamic potentials and their thermodynamic relations in nonextensive thermodynamics are the same as those in the classical thermodynamics [7, 8].

In this paper, under the framework of nonextensive thermodynamics, we study the generalized Gibbs free energy $G_q$ in section 2, the heat capacity at constant volume $C_{Vq}$ and heat capacity at constant pressure $C_{Pq}$ in section 3, and the generalized enthalpy $H_q$ in Sec.4. The conclusion is given in section 5.

## 2. On the Generalized Gibbs free energy $G_q$

Tsallis q-entropy in nonextensive thermodynamics is nonextensive [1]. Abe studied the thermodynamic potential by generalized zeroth law of thermodynamics and the so-called physical temperature $T_{phys}$ [8], which is not the inverse of the Lagrange multiplier $\beta$, defined as

$$T_{phys} = \left(1 + \frac{1-q}{k_T} S_q \right) \frac{1}{k_T \beta} \tag{1}$$

where $k_T$ is a constant, which may depend on nonextensive parameter $q$, and returns to the Boltzmann constant $k_B$ in the limit $q \to 1$. The generalized Helmholtz free energy $F_q$ is given by

$$F_q = U_q - T_{phys} \frac{k_T}{1-q} \ln\left(1 + \frac{1-q}{k_T} S_q \right), \tag{2}$$

which is different from traditional thermodynamics. Therefore, we can define the generalized Gibbs free energy $G_q$ as

$$G_q \equiv F_q + P_{phys} V, \tag{3}$$

where $P_{phys}$ is the physical pressure [8] defined by

$$P_{phys} = \frac{T_{phys}}{1 + \frac{1-q}{k_T} S_q} \frac{\partial S_q}{\partial V} \tag{4}$$

Then,

$$G_q = U_q - T_{phys} \frac{k_T}{1-q} \ln\left(1 + \frac{1-q}{k_T} S_q\right) + P_{phys} V \tag{5}$$

and

$$dG_q = dU_q - \frac{k_T}{1-q} \ln\left(1 + \frac{1-q}{k_T} S_q\right) dT_{phys} - T_{phys} \frac{dS_q}{1 + \frac{1-q}{k_T} S_q} + P_{phys} dV + V dP_{phys} \tag{6}$$

Using the first law of thermodynamics,

$$d'Q_q = dU_q + P_{phys} dV, \tag{7}$$

where $Q_q$ is the quantity of heat, satisfying

$$dS_q = \left(1 + \frac{1-q}{k_T} S_q\right) \frac{d'Q_q}{T_{phys}}, \tag{8}$$

one has

$$dU_q = T_{phys}\left(1 + \frac{1-q}{k_T} S_q\right) dS_q - P_{phys} dV \tag{9}$$

Substitute Eq.(9) into Eq.(6), it becomes

$$dG_q = -\frac{k_T}{1-q} \ln\left(1 + \frac{1-q}{k_T} S_q\right) dT_{phys} + V dP_{phys} \tag{10}$$

and therefore,

$$\left(\frac{\partial G_q}{\partial T_{phys}}\right)_{P_{phys}} = -\frac{k_T}{1-q} \ln\left(1 + \frac{1-q}{k_T} S_q\right) \tag{11}$$

$$\left(\frac{\partial G_q}{\partial P_{phys}}\right)_{T_{phys}} = V \tag{12}$$

Eq.(11) is different from that in the traditional thermodynamics, while Eq.(12) is the same and so invariant in the framework of nonextensive thermodynamics. Because the physical temperature is not the inverse of the Lagrange multiplier, some of the relations between the thermodynamic potentials are changed, while some are not.

## 3. The thermodynamic relation between $C_{Vq}$ and $C_{Pq}$

In the classical thermodynamics, we know there is an important thermodynamic relation between the heat capacity at constant volume, $C_V$, and the heat capacity at constant pressure, $C_p$. Namely [9],

$$C_p - C_V = -T \left(\frac{\partial V}{\partial T}\right)_p^2 \Big/ \left(\frac{\partial V}{\partial p}\right)_T. \qquad (13)$$

We study this thermodynamic relation on $C_{Vq}$ and $C_{Pq}$ in nonextensive thermodynamics. According to the definition of heat capacity at constant volume $C_{Vq}$ and at constant pressure $C_{Pq}$, respectively,

$$C_{Vq} = \left(\frac{\partial Q_q}{\partial T_{phys}}\right)_V, \qquad (14)$$

$$C_{Pq} = \left(\frac{\partial Q_q}{\partial T_{phys}}\right)_{P_{phys}}, \qquad (15)$$

and using Eq.(8), one has

$$C_{Vq} = T_{phys}\left(1 + \frac{1-q}{k_T} S_q\right)\left(\frac{\partial S_q}{\partial T_{phys}}\right)_V, \qquad (16)$$

$$C_{Pq} = T_{phys}\left(1 + \frac{1-q}{k_T} S_q\right)\left(\frac{\partial S_q}{\partial T_{phys}}\right)_{P_{phys}}, \qquad (17)$$

where the derivative of Tsallis entropy, in terms of the Jacobian determinant and its property, is

$$\left(\frac{\partial S_q}{\partial T_{phys}}\right)_V = \frac{\partial(S_q, V)}{\partial(T_{phys}, V)} = \frac{\dfrac{\partial(S_q, V)}{\partial(T_{phys}, P_{phys})}}{\dfrac{\partial(T_{phys}, V)}{\partial(T_{phys}, P_{phys})}}$$

$$= \left[\left(\frac{\partial S_q}{\partial T_{phys}}\right)_{P_{phys}}\left(\frac{\partial V}{\partial P_{phys}}\right)_{T_{phys}} - \left(\frac{\partial S_q}{\partial P_{phys}}\right)_{T_{phys}}\left(\frac{\partial V}{\partial T_{phys}}\right)_{P_{phys}}\right] \Big/ \left(\frac{\partial V}{\partial P_{phys}}\right)_{T_{phys}} \qquad (18)$$

Substitute Eq.(18) into Eq.(16), one obtains

$$C_{Vq} = T_{phys}\left(1+\frac{1-q}{k_T}S_q\right)\left(\frac{\partial S_q}{\partial T_{phys}}\right)_{P_{phys}} - T_{phys}\left(1+\frac{1-q}{k_T}S_q\right)\left(\frac{\partial S_q}{\partial P_{phys}}\right)_{T_{phys}}\left(\frac{\partial V}{\partial T_{phys}}\right)_{P_{phys}} \bigg/ \left(\frac{\partial V}{\partial P_{phys}}\right)_{T_{phys}}$$

. (19)

Using Eq.(11) and (12), one gets

$$\frac{\partial^2 G_q}{\partial P_{phys}\partial T_{phys}} = -\frac{1}{1+\frac{1-q}{k_T}S_q}\left(\frac{\partial S_q}{\partial P_{phys}}\right)_{T_{phys}}, \quad (20)$$

and

$$\frac{\partial^2 G_q}{\partial T_{phys}\partial P_{phys}} = \left(\frac{\partial V}{\partial T_{phys}}\right)_{P_{phys}}. \quad (21)$$

So one finds

$$-\frac{1}{1+\frac{1-q}{k_T}S_q}\left(\frac{\partial S_q}{\partial P_{phys}}\right)_{T_{phys}} = \left(\frac{\partial V}{\partial T_{phys}}\right)_{P_{phys}}. \quad (22)$$

Substitute Eq.(22) into Eq.(19), we derive the relation between $C_{Vq}$ and $C_{Pq}$ in nonextensive thermodynamics,

$$C_{Pq} - C_{Vq} = -T_{phys}\left(1+\frac{1-q}{k_T}S_q\right)^2\left[\left(\frac{\partial V}{\partial T_{phys}}\right)_{P_{phys}}\right]^2 \bigg/ \left(\frac{\partial V}{\partial P_{phys}}\right)_{T_{phys}}. \quad (23)$$

Clearly, this relation for the heat capacity has a different form from the classical thermodynamic relation (14) if the temperature is defined in terms of the physical temperature in nonextensive thermodynamics; only if one takes $q=1$, it becomes the classical form Eq.(13).

**4. On the Generalized enthalpy $H_q$**

In traditional classical thermodynamics, there are fundamental thermodynamic relations between the thermodynamical potentials and heat capacity. Firstly, in nonextensive thermodynamics, the generalized enthalpy $H_q$ is defined as

$$H_q \equiv U_q + P_{phys}V. \quad (24)$$

Therefore, the relation between the thermodynamical potentials is form invariant in the framework of nonextensive thermodynamics,

$$G_q - F_q = H_q - U_q. \tag{25}$$

From Eq.(24), one gets

$$dH_q = dU_q + P_{phys}dV + VdP_{phys}. \tag{26}$$

Using Eq.(9), it becomes

$$dH_q = T_{phys}\left(1 + \frac{1-q}{k_T}S_q\right)dS_q + VdP_{phys}. \tag{27}$$

So

$$\left(\frac{\partial H_q}{\partial T_{phys}}\right)_{P_{phys}} = T_{phys}\left(1 + \frac{1-q}{k_T}S_q\right)\left(\frac{\partial S_q}{\partial T_{phys}}\right)_{P_{phys}} \tag{28}$$

Comparing with Eq.(17) one finds

$$\left(\frac{\partial H_q}{\partial T_{phys}}\right)_{P_{phys}} = C_{Pq}, \tag{29}$$

and with Eq. (16), one obtains

$$\left(\frac{\partial U_q}{\partial T_{phys}}\right)_{V_{phys}} = C_{Vq}. \tag{30}$$

They are form invariant if the temperature is defined using the physical temperature.

## 5. Conclusion

In this paper, we derive the generalized Gibbs free energy and the generalized enthalpy in the framework of nonextensive thermodynamics by using the physical temperature and the physical pressure. We study the generalized thermodynamical potetials, the generalized Gibbs free energy $G_q$, the generalized Helmholtz free energy $F_q$, the generalized enthalpy $H_q$, and their relations. We also study an important thermodynamical relation between the heat capacity at constant volume, $C_V$, and the heat capacity at constant pressure, $C_p$ in nonextensive framework.

We derive the important relation in the classical thermodynamics between the heat capacities at a constant volume and at a constant pressure, Eq.(23), by using the generalized thermodynamical potential. We find it has a different form from the traditional one in Gibbs thermodynamics. But, the expressions of the heat capacities using the generalized thermodynamical potentials, Eq(29), Eq.(30), in this framework are still the same as the traditional one. Therefore, the difference between the physical temperature and the inverse of Lagrange multiplier can't affect the forms of the heat capacities.

**Acknowledgement**

This work is supported by the National Natural Science Foundation of China under grant No.10675088.